\documentclass[aps,pra,floatfix,twocolumn]{revtex4}
\usepackage{amsmath}
\usepackage{amssymb}
\usepackage{epsfig}
\usepackage{subfigure}
\usepackage{graphicx}
\usepackage{units}
\usepackage{epstopdf}
\usepackage{color}

\newcommand{\ket}[1]{\vert#1\rangle}

\def\Tr{\hbox{Tr}} 
\begin{document}
\title{Non-Gaussianity of quantum states: an experimental test on
single-photon added coherent states} 
\author{Marco Barbieri$^1$, Nicol\`o Spagnolo$^{1,2}$, Marco G. Genoni$^{3}$,  Franck  Ferreyrol$^1$, \\R\'emi  Blandino$^1$, Matteo G.A. Paris$^{3}$, Philippe Grangier$^1$, and Rosa Tualle-Brouri$^1$} 
\affiliation{$^1$ Laboratoire Charles Fabry , Institut d'Optique, 
Universit\'e Paris-Sud and CNRS, F-91127, Palaiseau, France}
\affiliation{$^2$ Dipartimento di Fisica, ``Sapienza'' Universit\`a di 
Roma and CNISM, I-00185, Roma, Italy}
\affiliation{$^3$ Dipartimento di Fisica dell'Universit\`a degli Studi di Milano}
\affiliation{CNISM, UdR Milano Universit\`a, I-20133 Milano, Italy}

\begin{abstract}
Non Gaussian states and processes are useful resources in quantum information with continuous variables. 
An experimentally accessible criterion has been proposed to measure the degree of non Gaussianity of quantum states, based on
the conditional entropy of the state with a Gaussian reference. Here we adopt such criterion to characterise an important class of non classical 
states, single-photon added coherent states. Our studies demonstrate the reliability and sensitivity of this measure, and use it to quantify
how detrimental is the role of experimental imperfections in our realisation.
 
\end{abstract}

\maketitle

Quantum information offers a different viewpoint on fundamental aspects
of quantum mechanics: it aims to assess and exploit the quantum 
properties of a physical system as a resource for different, and 
wishfully more efficient, treatment of information. Indeed, within the 
framework of quantum information with continuous variables \cite{Bra05}, 
nonclassical states of the radiation field represent a resource and 
much attention has been devoted to their generation schemes, which usually 
involves nonlinear interaction in optically active media. 

On the other hand, the reduction postulate provides an alternative mechanism to achieve 
effective nonlinear dynamics;  if a measurement is performed on 
a portion of a composite entangled system, the other component is 
conditionally reduced according to the outcome of the measurement. The 
resulting dynamics may be highly nonlinear, and may produce quantum 
states that cannot be generated by currently achievable nonlinear 
processes. Conditional measurements have been exploited to engineer 
nonclassical states and, in particular, have been recently employed 
to obtain non-Gaussian states. 

While Gaussian states, defined as those states with a Gaussian Wigner function, are known 
to provide useful resources for tasks such as teleportation \cite{Bra98,Fur98},
cloning \cite{Bra01,Coc04,And05}, or dense coding \cite{Ban99,
Bra00,Li02}, there is an ongoing effort to study which protocols are
allowed by non-Gaussian resources. The most notable example is certainly
their use for an optical quantum computer \cite{Ral03, Lun08}, alongside
with their employment for improving teleportation \cite{Opa00,Coc02,Oli03}, 
cloning \cite{Cer05}, and storage \cite{Cas07}. Several realisations of non-Gaussian states have
been reported so far, in particular from squeezed light
\cite{Lvo01,Wen04,Zav04,Our06,Our06a,Nee07,Our07a,Kim08,Our09}, close-to-threshold parametric oscillators \cite{Dau05,Dau10}, in
optical cavities \cite{Del08}, and in superconducting circuits
\cite{Hof08}. Non-Gaussian operations are also interesting for tasks as
entanglement distillation \cite{Our07,Jap10}, and noiseless
amplification \cite{Fer10,Xia10} which also are obtained in a
conditional fashion, accepting only those events heralded by a
measurement result.
\par
In principle, non-Gaussianity is not directly related to the non-classical 
character of a quantum state and, in turn, classical non-Gaussian 
state may be prepared, e.g. by phase-diffusion of coherent states
or photon subtraction on thermal states \cite{All10}.
On the other hand, in the applications mentioned above it is the
presence of both non-Gaussianity and non-Classicality which allows for
enhancement of performances.  Therefore, de-Gaussification 
protocols of interest for quantum information are those providing 
non-Gaussianity in conjunction with nonClassicality. 

\begin{figure}[b]
\includegraphics[viewport = 30 310 550 500,clip, width=\columnwidth]{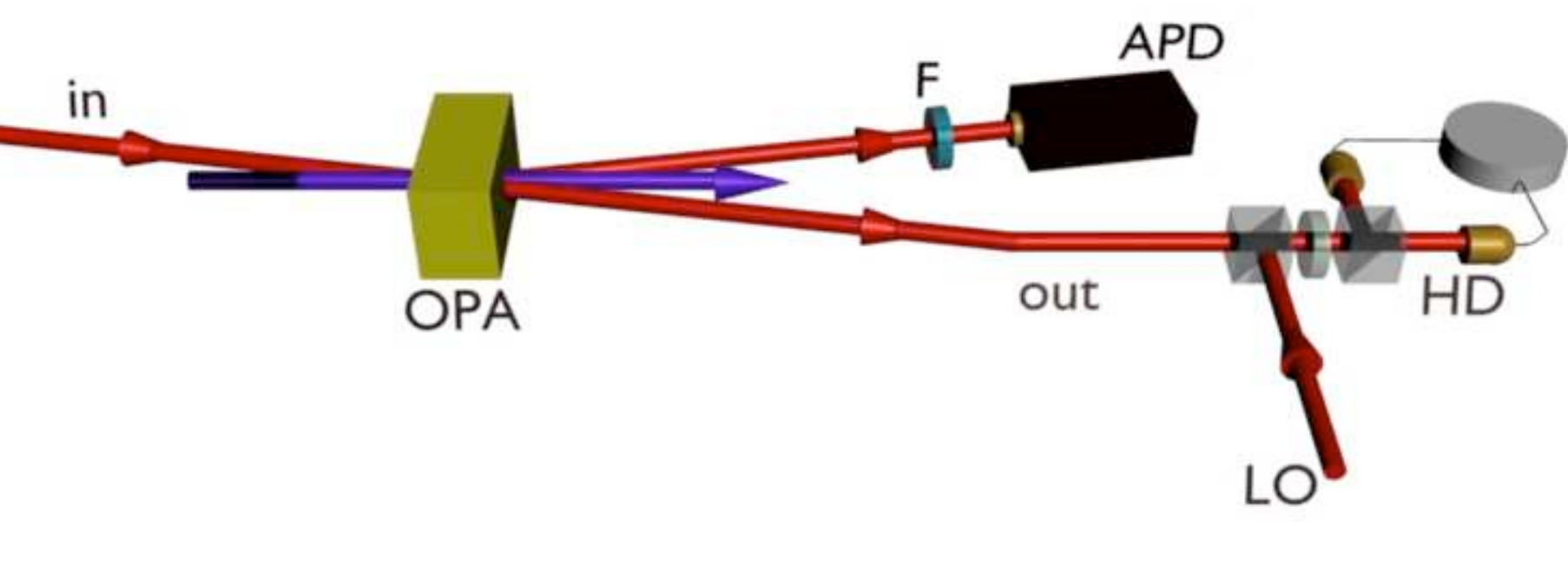}
\caption{Layout of the experiment. An optical parametric amplifier (OPA)
is injected with a coherent state of variable amplitude $\|\alpha\|$ in
the range $[0,1.5]$. This is realised by a $100\mu m$ thick slab of
potassium niobate, pumped by a frequency-doubled mode-locked Ti:Sapphire laser
($\lambda_p{=}425$nm, pulse duration $230$fs, repetition rate 800kHz).
Our OPA is driven in frequency-degenerate and non-collinear
regime, so to generate an idler  at the same wavelength
$\lambda=2\lambda_p$ as the coherent seed; this is then spatially
filtered with a single-mode fibre, spectrally filtered by a diffraction
grating and a slit, indicated as F in the figure. Finally, the idler is detected by an avalanche photodiode APD. The
observation of the output conditioned by an APD count results in
single-photon addition. The quantum state of the output is reconstructed
by homodyne detection (HD). Mode-matching with the local oscillator (LO)
employs polarisation: the signal and the LO are first matched on a
polarisation beam splitter, and then combined using a half-wave plate
and a second polariser so to realise an accurate 50:50 intensity
splitting.} \end{figure} 

In this work we address the conditional dynamics induced by the
so-called photon addition as a protocol to generate nonclassical 
non-Gaussian states. We quantify experimentally the amount of 
non-Gaussianity obtained by adding a photon to a coherent state \cite{Zav04,Zav07a,Par07,Zav09}.
Differently from previous investigations \cite{Shc05,Zav07a,Kie08,Kie09,Spa09}, we can explicitly 
address the two aspects of non-Gaussianity and non-Classicality at once. 
For the former, we adopt the non-Gaussianity measure $\delta[\varrho]$ proposed in
\cite{Gen08,Gen10}, defined as the quantum relative entropy between the
quantum state itself $\varrho$ and a reference Gaussian state $\tau$ having the
same covariance matrix as $\varrho$. Given this choice of the reference
Gaussian state, we have that $\Tr[\varrho \log \tau] = \Tr[\tau \log \tau]$, as $\log \tau$ is a polynomial of order at most 
two in the canonical variables \cite{Gen08,Holevo}. We thus find
\begin{align}
\label{delta}
\delta[\varrho] &= \cal S(\varrho \rVert \tau) = 
\Tr [\varrho (\log \varrho -\log \tau)] \nonumber \\
&=\cal S(\tau) - \cal S(\varrho)
\end{align}
that is, $\delta[\varrho]$ is simply equal to the difference between the von
Neumann entropy of $\tau$ and the von Neumann entropy of $\varrho$. In
Ref. \cite{Gen08} it has been shown that this measure is non zero only
for non-Gaussian states. It is also additive under tensor product, 
invariant under unitary Gaussian operations, and in general it does not
increase under generic completely positive Gaussian channels. This
measure is somehow preferable to that based on the Hilbert-Schmidt
distance \cite{Gen07} in a quantum information context, since it is
based on an information-related quantity. 
We note, however, that a mixture (e.g. doubly peaked) of classical states can also be strongly non-Gaussian: we therefore adopt an additional "non-classicality " criterion.

Several measures of non-classicality have been proposed in literature
\cite{Dod00,Dod03,Mar04,eisert}, for our purposes we consider as a witness a 
quantity $\nu[\varrho]$ related to the negativity of the Wigner function. This is normalised to a reference, which we choose to be a single photon state $W_1(x,p)$.
The non-classicality is then defined as 
\begin{align}
\nu[\varrho]= \frac{ \min\left(W(x,p)\right)}{ \min\left(W_1(x,p)\right)}.
\end{align}
This reference has been chosen since it has the lowest value within the class 
of states we consider. While this does not constitute a measure, it acts as a witness for non-Classical states whenever ${\nu[\varrho]>0}$.
The choice of using the single-photon as a reference is dictated by the need of a measure which does not depend on the convention for the quadratures. 
Moreover, it sets to unity the highest value of $\nu[\varrho]$ attainable in the class of states under investigation.
\par
A conceptual scheme of the experiment is shown in Fig. 1: an input
coherent beam $\ket{\alpha}$ is injected in an optical parametric
amplifier (OPA). This is a three-wave nonlinear interaction between a
pump beam and the input beam (usually called the signal $s$) which
results in the generation of a third beam called idler ($i$). When the pump
is an intense beam, we can treat it as a classical field: the output
state of the $s$ and $i$ modes can then be expressed as the application
of the squeezing operator: 
\begin{equation}
\label{sq}
S_{s,i}(r){=}\exp\left(r(a_s^\dag a_i^\dag-a_i a_s)\right),
\end{equation}
to the input $\ket{\alpha}_s\ket{0}_i$. Here, $r$ is the
squeezing parameter, which depends on the pump intensity, the crystal
length and its non-linear coefficients, and attained a value
${r{\simeq}0.105}$ in our experiment; we can then approximate
$S_{s,i}(r)$ taking the limit of weak nonlinearity: 
\begin{equation}
\label{eq:weakS}
S_{s,i}(r){\simeq}I+r(a_s^\dag a_i^\dag)-r(a_i a_s).  
\end{equation}


\begin{figure}[b!]
\includegraphics[width=.9\columnwidth]{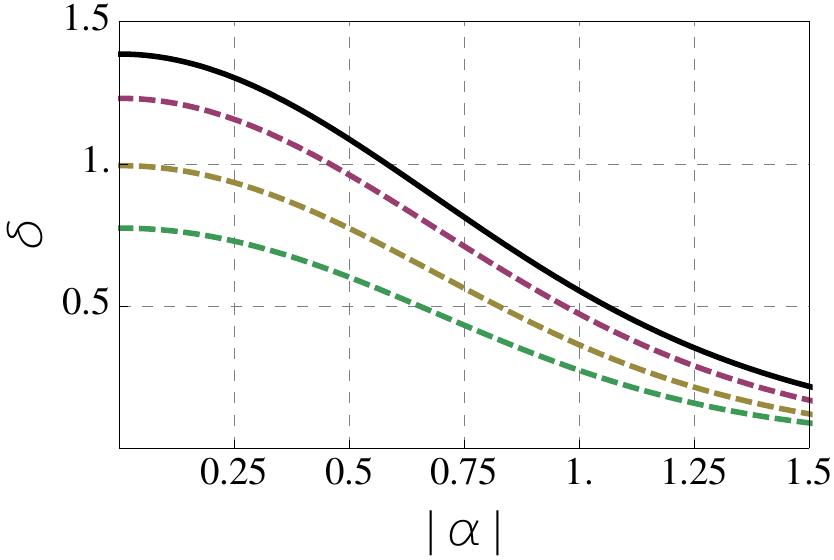}
\includegraphics[width=.9\columnwidth]{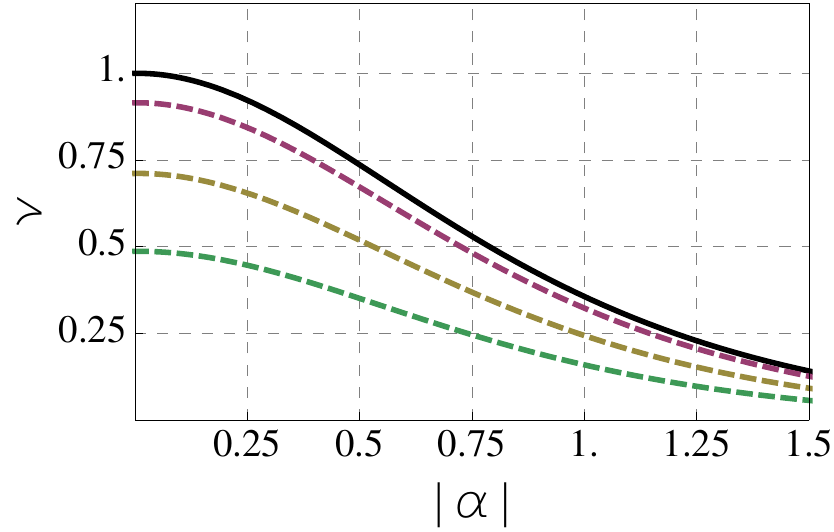}
\caption{\label{f:nGsqueezing} Non-Gaussianity $\delta[\varrho]$ (upper panel) and
non-classicality $\nu[\varrho]$ (lower panel) as 
a function of the amplitude $| \alpha |$ of the input coherent state for different
values of the squeezing parameter $r$ (dashed lines); from top to bottom 
$r = \{0.15,0.30,0.45\}$. The black solid line corresponds
to the non-Gaussianity of the ideal photon added coherent state, that is 
to the limit $r\rightarrow 0$.}
\end{figure}



\begin{figure*}[t] \includegraphics[width=1.732\columnwidth, viewport = 100 150 1000 500, clip]{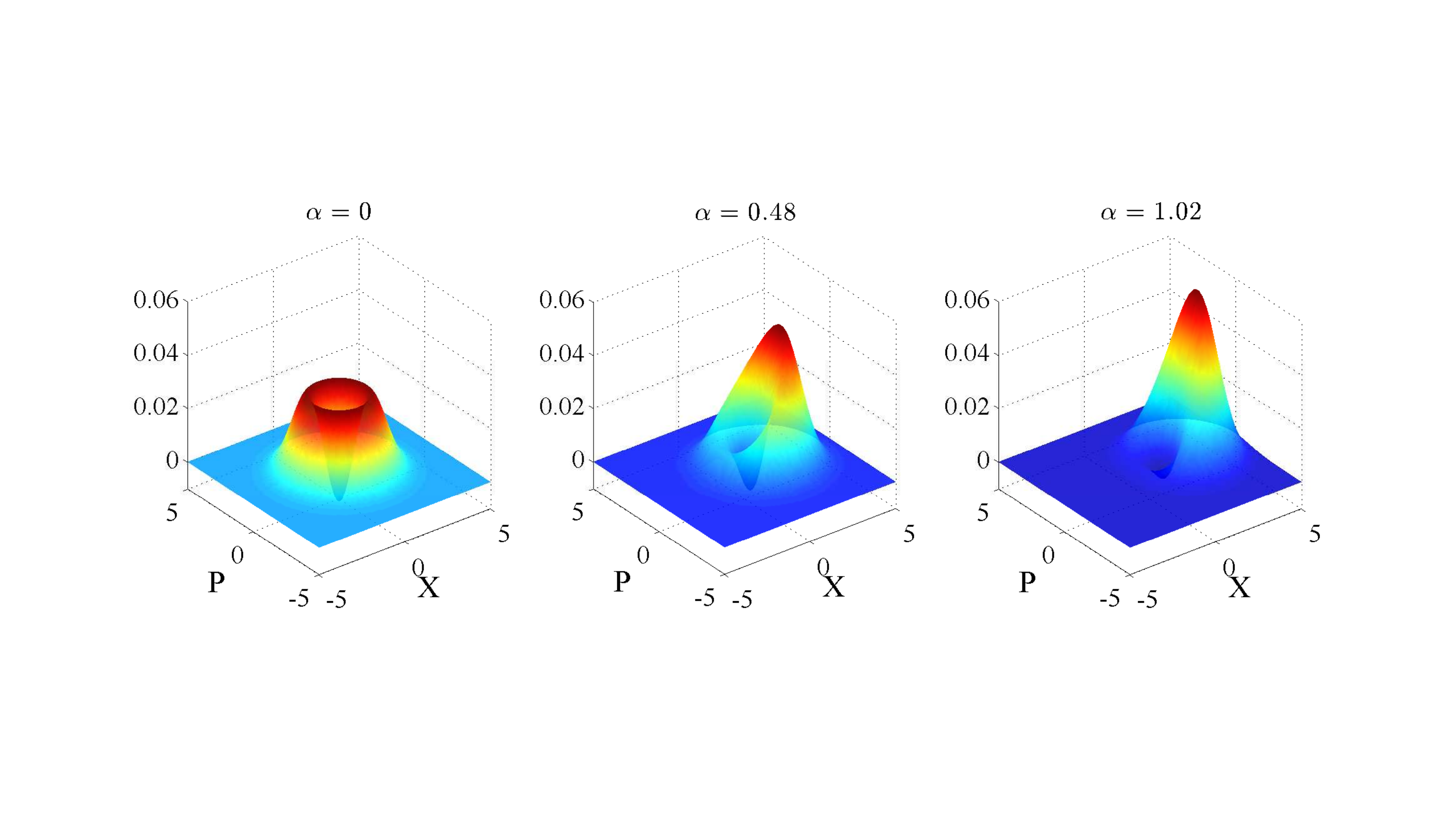}
\caption{\label{f:Wigner}  Experimental Wigner functions for increasing values of $\alpha$. The output states are reconstructed by a maximal likelihood
algorithm \cite{Lvo04} which interpolates 800,000 data points sorted
according to their phase into 12 histograms. The effectiveness of the sorting algorithm sets a lower bound $|\alpha|\sim0.5$, so that the oscillations due to interference are much larger than low-frequency noise fluctuations. Notice that for $\|\alpha\|{=}0$ this noise can be compensated by using a moving average technique.}
\end{figure*}


We now put an avalanche photodiode (APD) on the idler beam, and accept
only those events when a click is registered. Since the idler was
originally in the vacuum state, the only term which can give a
contribution in $S_{s,i}(r)$ is the second one. Therefore, the detection
of a single photon on the idler heralds the addition of a single photon
to the coherent state, transforming it into 
\begin{equation}
\frac{1}{\sqrt{1+\|\alpha\|^2}}a^\dag\ket{\alpha}, \label{eq:ideal}
\end{equation}
in the ideal case. In practice, we need a careful analysis of
those processes which spoil the photon addition and the non-Gaussianity of the
resulting state. Here, we follow closely the model presented in
Refs.\cite{Our06,Our07a,Our07}.

The detection on the idler beam is performed by an APD that can not
resolve photon number. In the limit of small detection efficiency, we
can approximate the detection process as the application of the $a_i$
annihilation operator on the idler mode. This is actually the case in
our experiment, where the overall detection efficiency is less than
10\%, due to spatial filtering ($\lesssim$75\%), spectral filtering
($\lesssim$30\%), and limited efficiency of the photodiode (55\%).

In Fig. \ref{f:nGsqueezing} we plot $\delta[\varrho]$ and $\nu[\varrho]$ 
as a function of the coherent amplitude $\alpha$, for different values of $r$. 
We observe that the two trends resemble closely, suggesting that 
the non-Gaussianity induced by photon-addition is essentially non-classical 
and thus useful for quantum information processing.
It can be also observed how both non-Gaussianity and non-classicality 
decrease by increasing the squeezing parameter;
this can be explained by observing that, as shown in
Eq.~(\ref{eq:weakS}), for low values of $r$, the squeezing operator adds
only one photon on each arm, while by increasing $r$ we have to consider
also the possible addition  of many photons. Due to the lack of photon
number resolution, the detection will be affected by the presence of
higher-number emission from the squeezing process, eq.~\eqref{sq}. In
this case, conditioned on a click from the idler beam, the signal will
be in a highly mixed and thus less non-Gaussian and also non-classical state. In the
\emph{ideal} limit of $r\rightarrow 0$ the non-Gaussianity of the state is exactly
equal to the one of the ideal photon added coherent state in
Eq.~(\ref{eq:ideal}). However, this goes to the expenses of the success
rate, and a compromise between non-Gaussianity and count rate has to be found.

Beside the role played by the squeezing, we have to take into account
the other imperfections that are present in our experimental setup.
In the OPA there might occur a certain modal mismatch between the pump and the input: this results in a parasitic amplification that introduces excess noise on the signal and idler modes. The process is modelled as a non-degenerate  OPA driven at a weaker strength $\gamma r$, where 
$0\leq \gamma \leq 1$ and $\gamma\sim 0.425$ in our experiment. The amplification couples the modes $s$ and $i$ with two other modes $s'$ and $i'$, initially in the vacuum state. The complete description takes the form:
\begin{equation}
S_{s,i}(r)S_{s,i'}(\gamma r)S_{s',i}(\gamma r)\ket{\alpha}_s\ket{0}_i\ket{0}_{s'}\ket{0}_{i'}.
\end{equation}
The parasite modes $s'$, $i'$ are not observed in the experiment, therefore we have to trace over them to obtain the output density matrix. 


\begin{figure}[b!] \includegraphics[width=.8\columnwidth]{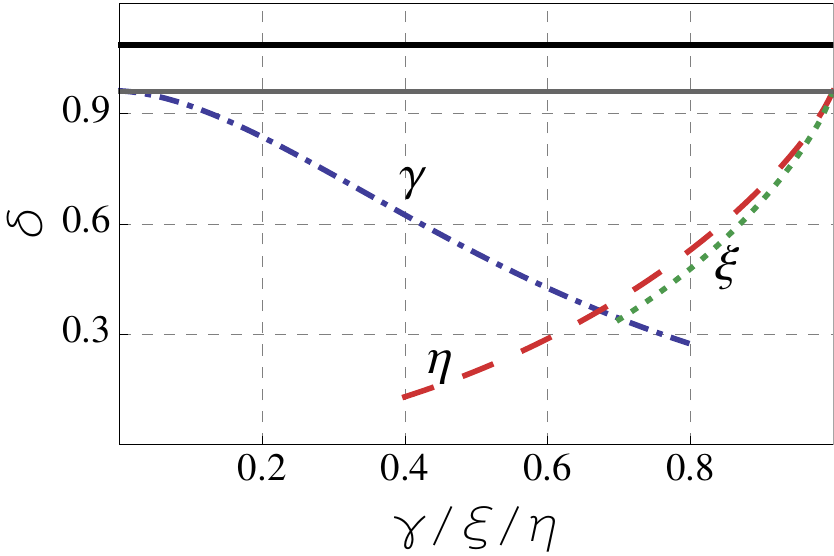}
\caption{\label{f:noise} Non-Gaussianity $\delta[\varrho]$ as a
function of the noise parameters of the experimental setup for fixed amplitude
$|\alpha|=0.5$ and squeezing parameter $r=0.15$. In details:  the blue dot-dashed
line corresponds to $\delta[\varrho]$ as a function of $\gamma$, the green dotted line 
as a function of $\xi$ and the red dashed line 
as a function of $\eta$. The solid lines refer to the non-Gaussianity of the ideal photon-added
coherent state with $|\alpha |=0.5$ (upper black line), and to the non-Gaussianity of the state obtained
by considering $|\alpha |=0.5$, squeezing parameter $r=0.15$ and no 
imperfections (lower grey line).}
\end{figure}


\begin{figure}[t!] 
\includegraphics[width=.75\columnwidth]{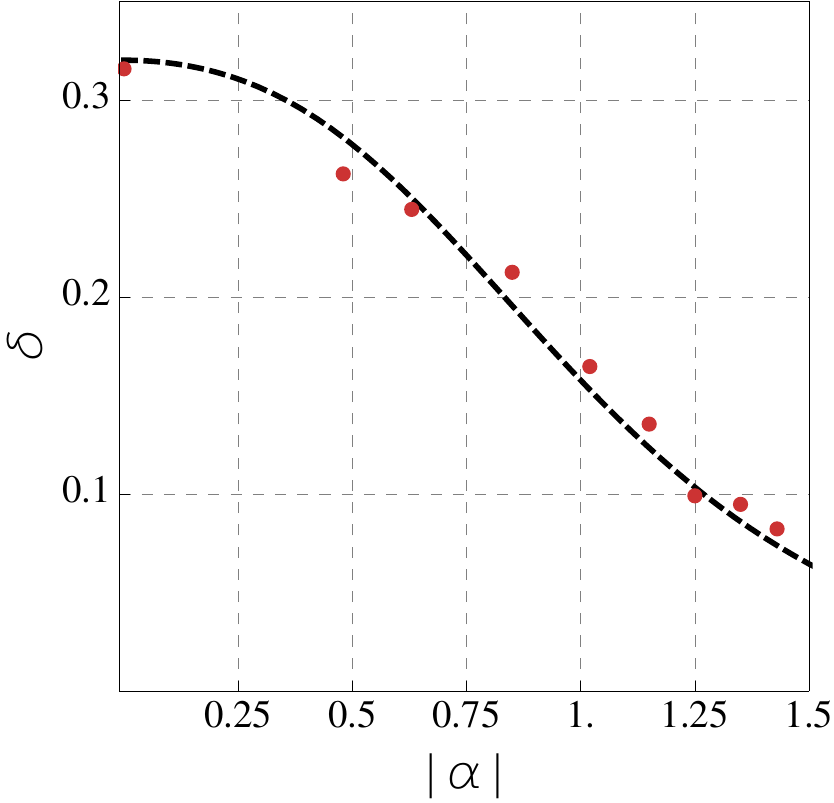}\\
\vskip .5cm
\includegraphics[width=.75\columnwidth]{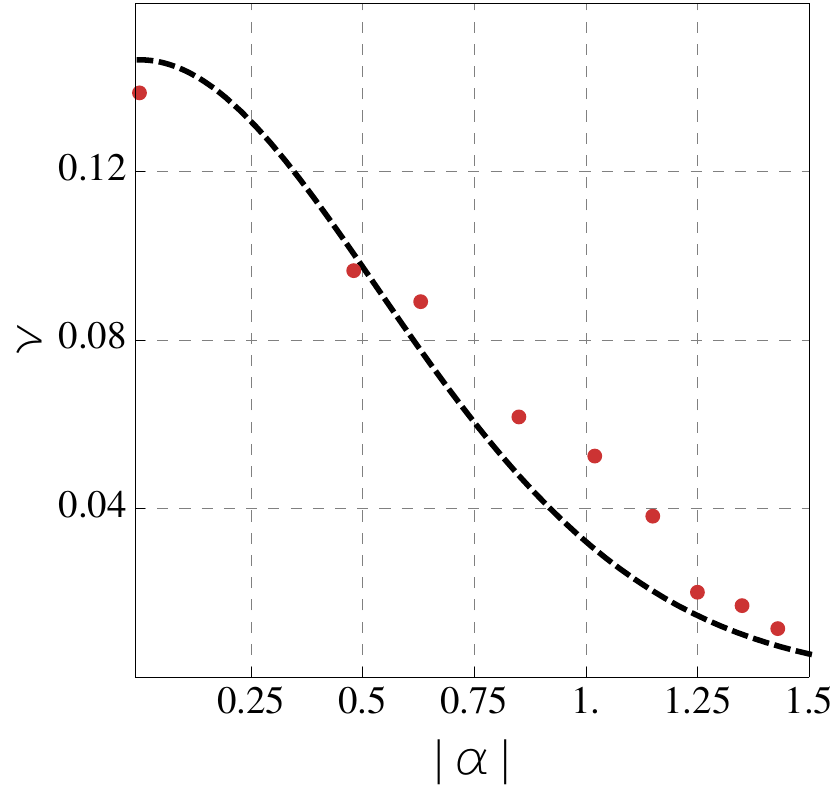}
\caption{\label{f:nonGExp} (Top): Non-Gaussianity $\delta[\varrho]$ as a
function of the amplitude $| \alpha|$ of the input coherent state. 
(Bottom):  non-classicality $\nu[\varrho]$  -- related to
the minimum value of the Wigner function of $\varrho$ -- as a function of $\alpha$.
The red points are the experimental values from the reconstructed matrices. 
The black dashed line is obtained from our model including the main experimental imperfections of our realisation. The parameters are chosen in such a way to fit the data of the non-Gaussianity: $r=0.105$, $\gamma=0.425 $, $\xi=0.96 $ and $\eta =0.71 $.}
\end{figure}


Accurate spatial and spectral filtering is performed so that the mode detected by the APD is matched with the input mode, that we detect by balanced homodyne; nevertheless, this task can be accomplished only with a limited efficiency $\xi$ which in our setup takes the value $\xi\sim0.96$. In formulae, we will have an output state $\varrho_{s,\checkmark}$ on the signal mode when the trigger count came from the correct mode, and a different state $\varrho_{s,{\rm x}}$ heralded by a faulty trigger event; the overall state is: 
\begin{equation}
\xi \varrho_{s,\checkmark}+(1-\xi)\varrho_{s,{\rm x}}.
\end{equation} 
Notice that dark count rates from the APD play a negligible role ($\sim10$ counts/s with an overall rate $\sim 1{-}4\cdot10^3$ counts/s), as we used a gated detection triggered by the cavity dumping electronics of our laser.
Homodyne detection has a limited efficiency as well, coming from optical loss, non-unit detector yield, and mode-matching between the local oscillator and the signal. The overall efficiency is, 
in our case, of the order of $\eta\sim0.71$; this is modelled as transmission through a beam splitter with transmittivity $t^2{=}\eta$. Examples of the measured Wigner quasi-distributions are illustrated in Fig.~\ref{f:Wigner}

In Fig.~\ref{f:noise} we plot  $\delta[\varrho]$ at fixed
values of the coherent state amplitude $\alpha=0.5$ and of the squeezing
parameter $r=0.15$ as a function, respectively, of the noise parameters $\gamma$, $\xi$ 
and $\eta$ chosen in ranges relevant for our experimental setup. 
We observe as expected that $\delta[\varrho]$ decreases monotonically 
with $\gamma$, while it increases monotonically with $\xi$ and $\eta$. 
For the values that characterize our experiment, the homodyne efficiency $\eta$ is the source of imperfection that affects in the most detrimental way the non-Gaussianity of our states.

Finally we evaluated $\delta[\varrho]$ from the experimentally
reconstructed output states for different coherent state amplitudes $|
\alpha |$; the results are shown in the upper panel of Fig.~\ref{f:nonGExp} . The
dashed line shows the description provided by our model when taking into account all
the \emph{noise} processes described above. The values of parameters used for the curve 
are obtained from a fit of the experimental data: $r=0.105$, $\gamma=0.425$,
$\xi=0.96$ and $\eta=0.71$. The average fidelity between the
reconstructed and the modelled state is $0.989\pm0.006$ \cite{noteonF}. Concerning the
non-Gaussianity, the agreement between the experimental data and the
model is satisfactory, and we can observe, as expected, a
decrease as the input intensity $|\alpha|$ increases.  As shown, the
effect of the single-photon addition is more relevant for quantum states
with a small number of photons, and becomes only a small perturbation
for  higher average photon number. 

In the lower panel of Fig.~\ref{f:nonGExp} we observe the behaviour
of $\nu[\varrho]$ as a function of the amplitude $\alpha$.
The experimental results confirm that the two quantities,
non-Gaussianity and non-classicality have a similar behaviour
and then that the non-Gaussianity induced by this photon-addition
operation is essentially non-classical.
As a general remark, we notice that the logarithmic term in the
expression \eqref{delta} amplifies the effect of small discrepancies
with the model we present. This qualifies our measure as a very
sensitive one in those contexts where a good estimation of information
resources is needed. In any case, our model is able to capture the
essential features of the process, and provides us a tool to quantify
how detrimental the imperfections are for the generation of these
non-Gaussian resources.


In summary, these experiments on single-photon added coherent states demonstrate the relevance of this recently proposed measure of the non-Gaussianity. This measure appears as a reliable and sensitive way to quantifying experimental imperfections of de-Gaussification experiments. It furthermore allows to exhibit a link between non-Gaussianity and non-Classicality in such experiments. More generally, it would
be useful to have at disposal a quantity able to capture both
features, non-Gaussianity together with non-classicality, for any
generic quantum states. Work along these lines is in progress and
results will be developed elsewhere.

We thank A. Ourjoumtsev, S. Olivares, F. Fuchs, F. Sciarrino, and F. De Martini for discussion.
This work is supported by the EU project COMPAS, the ANR SEQURE, and partially by the CNISM-CNR agreement. 
M.B. is supported by the Marie Curie contract PIEF-GA-2009-236345-PROMETEO.
F.F. is supported by CÕNano-Ile de France.




\begin{thebibliography}{99}
\bibitem{noteonF} Here, we use the definition of the fidelity $F$ between two states $\varrho_1$ and $\varrho_2$: $F{=}\Tr^2\left(\sqrt{\varrho_1}\varrho_2\sqrt{\varrho_1}\right)^{1/2}$.
\end{thebibliography}
\end{document}